# Tackling the "Gremlins Paradox":  Autonomous Hot Swap Healing Protocol for Chronic Performance Problem Aversion in On-board Mission Critical Systems on Lethal Autonomous Weapons [10]

by

Nyagudi Musandu Nyagudi – Security Analyst

Nairobi, Kenya – EAST AFRICA

**Abstract**

After careful reading of Kavulya [2013], [2] it was the informed view of this researcher, that the methods described in the Doctoral Dissertation could very easily resolve Chronic Performance Problems(CPPs) at Command Centers related to Lethal Autonomous Weapon Systems.  It is however not possible to use the same methods to resolve Chronic Performance Problems occurring in On-board Mission Critical Systems on Lethal Autonomous Weapons, that are unmanned and deployed/fielded on a non-interactive basis. Nyagudi [2014], [1] attempts to offer some insights, concepts, methods and approaches, that would facilitate the robust performance of lethal autonomous weapon systems.  With the benefit of insights from Kavulya [2014], [2] and a background in the physical sciences and knowledge in tactical operations realities, the researcher of this paper proposes an upgraded approach to Nyagudi [2014], [1] for delivering robust performance by way of a Autonomous Hot Swap Healing Protocol for Chronic Performance Problem Aversion in On-board Mission Critical Systems on Lethal Autonomous Weapons.  A major assumption for the actualization of this concept would be that micro-circuitry and nano-circuitry offer almost unlimited scope for hosting back-up systems to mission critical systems – without substantial increase of physical load that would generate resource challenges. But in the realm of physical sciences a paradox arises – the more miniaturization is achieved, the greater the likelihood for chronic performance problems in micro-circuitry and nano-circuitry due to externally induced and/or intrinsic parasitic phenomena both at quantum and electronic levels. Parasitic quantum phenomena and electromagnetic interference could be encountered in purportedly electronic circuits, due to miniaturization. A balance must be found that renders this **"Gremlins Paradox"** negligible.

## 0 Introduction

Article 36 of Additional Protocol I of the Geneva Convention [3] states that, "In the study, development, acquisition or adoption of a new weapon, means or method of warfare, a High Contracting Party is under an obligation to determine whether its employment would, in some or all circumstances, be prohibited by this Protocol or by any other rule of international law applicable to the High Contracting Party."

This article of international law obliges High Contracting parties to take all reasonable precaution in ensuring safety and compliance with international humanitarian law when deploying robot type lethal autonomous weapons, that can for all intents and purposes be classified as new and/or emerging weapons – the 1987 commentary [4] to Article 36 of Additional Protocol I of the Geneva Convention envisages High Contracting parties undertaking legal reviews pertaining to the development and deployment of new and emerging weapons.  Other norms of war such as reasonable precaution also necessitate legal reviews of new and/or emerging weapon system technologies.

If [1] is considered as an emerging technology/study, and some information has come to light via [2] that can improve its legality and performance – i.e. the researcher who drew up [1] has acquired more relevant knowledge by way of studying the concepts in [2] and other sources, he is therefore obliged by legal statute laid out in [3] and explained in [4] to upgrade the concept into legal compliance or to retract it all together, if he is unable to upgrade it.  In this case the researcher has determined that an upgrade is feasible for purposes of continuity and compliance, such that [1] is enabled to tackle CPPs.

## 1 Rationale

Miniaturization has enabled man to progress from low processing power multi-room size computers to high processing power hand held smart phones. The transistor has declined in size from vase sized tubes to micro-circuitry. With ever smaller transistor sizes, man has added computational power while reducing the sizes of the hosting systems. There is a limit to this phenomenon, a transistor is made up of thousands if not hundreds of atoms – it is therefore impossible to obtain a transistor that is smaller than an atom.

When circuits are reduced to molecular level, there is notable quantum interference, which makes it impossible to produce electronic circuitry using classical techniques. In the quest for the infinitesimally small electronic component to pack ever more computational capability into devices, man is faced with the challenges of parasitic electromagnetic, electrical and quantum interference, ie. The "Gremlins Paradox", due to the unintentional, counter-intuitive, counter-productive and problematic effects of the same, if classic electronic techniques alone are deployed.

Quantum interference[5] can be harnessed to create a new generation of electronic devices, but this is still just a theoretical possibility due to the fact that the molecules used are organic and the experimental set-ups are too delicate for real world usage in most instances. The occurrence of quantum interference in standard micro-circuitry would be parasitic in most instances and result in chronic performance problems, if the circuit in question is part of a computing device or facility. There are many other papers with similar content to those of [5]. It is very clear that quantum interference can cause unintended alteration to electronic bit strings.

Quantum level circuits would also be more sensitive to external electromagnetic and magnetic forces. Simply stated : Miniaturization allows for devices that increase the computational capacity of a system in ever more limited space, however it also causes intrinsic and deep rooted interference problems within the same circuits, that are likely to cause CPPs in a computing environment.

Since it is not possible to do away with current and emerging hardware, there should be a way of resolving the wide range of chronic performance problems. These may arise in circuits due to interference of various kinds at low level resulting in malfunction or under-performance of a system.

## 2 Reviews and Differentiation

Kopp [6] offers various photos and written explanations as to the mesh seen shrouding the Electro-optical seeker windows of various image homing missiles.  It is explained that this Faraday cage type feature shields the mission critical control systems therein from electromagnetic radio frequency induced interference [8].  The cumulative effect of exposure to the same may imply that a image homing missile may have disruption in its guidance and strike outside its targeted area.  Given that embedded and other computing systems are housed in missile warheads, this is an extreme but relevant case of the destruction that can be brought about via chronic performance problems.

Day to day experience with chronic performance problems in the personal computing environment are not strange, but may remain unclassified to the non-expert user.  At first, persons log on to their desktop, smart phone, tablet or laptop computers without much problem after booting up.  But after hours or days of usage, functions such as bluetooth become "stubborn" and multiple applications may not work as flawless as when the computer was first switched on – it is safe to assume then that this is the cumulative effect of chronic performance problems as detailed in [2].  This example puts into perspective the terminology of CPPs, for the most basic users of computing devices and systems.

When dealing with unmanned weapon systems, especially those that remain uncontrolled and autonomous for substantial periods of time, upon release into a battle-space chronic performance problems would be a major concern.  The range of problems that may arise would be, eg.:
1. performance variation : wrong re-targeting or faulty target identification
2. terminal malfunctions : loss of weapon
3. crippling malfunctions : weapons perform but do so outside desired time parameters

Previous theories that may easily explain such occurrences would have been those of Computational Complexity and Mathematical Chaos.  CPPs however deal with those and much more, making it difficult for a weapon systems engineer to properly estimate all possible outcomes, once a type of weapon is fielded.  These could be caused by intra-circuit heating, quantum interference, electromagnetic frequency interference, computational complexity, mathematical chaos, poor or faulty design, software bugs, etc.

These functional anomalies could affect navigation, targeting and economy of resource utilization within a lethal autonomous weapon system platform.  Unlike in the domain of commercial services, whose usage of production systems is all about profits and losses, in the realm of warfare lethal autonomous weapons and production systems based in command centers deal with matters of life and death, hence the importance of completely minimizing if not eradicating any possibility of CPPs.

Causes of CPPs could be:
1. Hardware design, damage or defect problems
2. Faulty configuration
3. Software design, damage or defect problems
4. etc.

Given the scope of these source of problems, there is no computer system running in the world today, without CPPs of one kind or another.  An extrapolation of this would be the fact that indeed all lethal autonomous weapon systems have some form of CPPs that have not been detected.  In the view of this researcher, it is an issue that has not been raised in any of the past research papers into lethal autonomous weapons.  Failure to provide a convincing method for dealing with the problem would render it illegal to use any kind of lethal autonomous weapon, especially of the robotic kind, that is unleashed into a battle-space as an autonomous warfighter.

A generic method for dealing with CPPs on-board lethal autonomous weapon systems is therefore proposed in this paper.  The methods proposed for use in Production Systems by [2] do not meet this requirement on an "as is" basis for the following reasons:
1. a human systems administrator is required
2. they are specific to particular hardware and software and are not general guidelines
3. for the systems in [2] their failure do not necessarily translate into life and death issues
4. many of the methods may discover but do not resolve the problems
5. CPPs may affect the tools that are used to discover them, complicating the unresolved "Gremlins Paradox"
6. their implementation by remote control leads to the possibility of hacking into a weapon system
7. Electromagnetic spectrum emissions may betray weapon location and concealment

But the methods in [2] for dealing with the problem of CPPs may work very well for highly complex military command centers, where data from many fielded weapon systems is converged and processed. The computational platforms therein are usually extensive and prone to faults and disruption of the kinds detailed in [2]. Lethal autonomous weapons in most cases are designed and fielded on a basis of necessity, i.e. communications can be interfered with by a skilled adversary, time-lapse between transmission and reception may be unacceptable in the specific domain of a mission critical system.

To add on to this mix, would be the fact that if methods of [2] are deployed remotely on a lethal autonomous weapon system, they may add more overhead to the OODA(observe, orient, decision, action) loop, exposing the platform and its mission to potential failure via adversarial countermeasure.

In the military procurement and logistics chain, issues pertaining to CPPs are not new, indeed the concept applies to all manner of components computers, electronics, mechanical, hydraulic, etc. [7] describes them as "lemons", when they are repeatedly "repaired" in a bid to make a fielded system to work as per expectation. [7] also states that the more complex a weapon system is, the more reduced its availability and the greater its maintenance costs. "Lemons" were viewed as troublesome in the military logistics chain and all efforts were made to identify them in the low production phase of a weapon system during field trials. Field trials are used to collect as much data as possible and all components have serial numbers, thereby making it easy to pick out the "lemons", by way of databases.

Data analysis leads to weapon system re-design and maturity, as a lot of the information is discovered via databases during field trials. It is not difficult to infer at this point that other causes of "lemons" may include the issue of low quality components due to counterfeits in the supply chain. At the turn of the century more and more computers were introduced into weapon systems. These increasingly complex devices came with many CPPs that are unlikely to be identified by way of field trials.

Presumably though field trials continue during the low production phases of weapon systems, but CPPs are not readily identified then and the "lemons" became more and more likely to show up in the naval, air and land battle-space. Since some weapon systems use commercially available off-the-shelf computers such as laptops, some of the CPPs experienced day to day by personal computer users are also experienced in the domain of mission critical computing in the military.

A review of a sample of the military grants/contracts may easily be obtained demonstrating that CPPs are a critical issue that have been identified by the Department of Defense of the Government of the United States of America(USA). Examples of these are the DARPA PCES contract F33615-03-C-4110 of [2], the Army Research Office, Grant Number DAAD 19-02-1-0389, "Perpetually Available and Secure Information Systems", to the Center for Computer and Communications Security at the Carnegie Mellon University of [9] and the United States Army, Contract No. MDA 903-91-C-0006 of [7]. A more detailed search of various digital libraries is likely to come up with similar funding towards the same class of problems.

[9] had some interesting insights but was highly dependent on the concept of fault injection. For the United States Army to have given a grant for a mission critical solution, then to almost immediately avail the research on the Internet, may be an indication that they did not at the time see immediate applicability of the research findings and were putting out feelers to the research community.

[2] also had the benefit of military funding and so did [7] before it, clearly demonstrating the felt need for solutions of various kinds within military circles in the USA as the problem became more evident. [9] describes the concept of white box techniques that are specific to obtaining performance data from applications where the source code is known, while black box techniques deal with gleaning information from hardware and operating system level where the specifics are shielded/proprietary.

Another class of techniques described as gray box that share characteristics of black box and white box techniques. To this researcher the problems encountered in [2] and [9] that necessitate black box techniques can easily be overcome by a system user that has got large purchasing and specification power such as the Department of Defense of the USA, that spends funds to the amount that it can demand for complete details, design and specification of each and every component it buys. In such a scenario, a researcher into the realm of CPPs may be in a position to know of the precise make up of a processor even if he or she is not in a position to interact with those functions directly – an additional scenario that we may classify in this paper as **glass box techniques**.

Converting the extensive amount of research in the domain of CPPs in computing systems/devices into usable knowledge must be a major preoccupation of the USA Military, its objectives may not be immediately met by academia due to sublime pursuits by the same or lack of exposure to some types of classified military system problems.  Perpetually Available and Securely Functioning Lethal Autonomous Weapons, must be within this category of problems that may only be disclosed to limited circles in academia, despite the fact that they are a matter of great interest to a global audience.

The "**Autonomous Hot Swap Healing Protocol for Chronic Performance Problem Aversion in On-board Mission Critical Systems on Lethal Autonomous  Weapons**" attempts to address these issues to a global audience, since weapon systems deployed must be understood by international war crimes tribunals, manufacturers, humanitarian organizations, warfighters, legislative bodies, etc.  Issues of applicability in the military domain may not be addressed adequately in [2] and [9] probably due to lack of exposure to military technology and problems, though it is not stated as such in those papers.

3 The Protocol

**Autonomous Hot Swap Healing Protocol for Chronic Performance Problem Aversion in On-board Mission Critical Systems on Lethal Autonomous  Weapons,** advocates for a virtual machine implementation of computing systems that are the mission critical components of lethal autonomous weapons.  A virtual machine is isolated from the hardware platform and can be specially written and tested to some level of specificity that eliminates the potential for easily avoidable CPPs.  The virtual machine would be hosted on a virtual machine platform, which in turn  interacts with the hardware platform directly.

The virtual machine platform would black list and avoid interacting with faulty hardware components, further reducing the likelihood of avoidable CPPs.  There is anecdotal evidence that this is possible.  [2 : page 111, Incident 4] states that the Hadoop has a mechanism for transfer of tasks from nodes that are affected by CPPs or other problems and it in turn black lists those faulty nodes.  Someone with experience of coding such a function within the Hadoop, could be of use when extrapolating the concept to other platforms such as the proposed class of virtual machine platforms for various lethal autonomous weapons.

These novel virtual machines could be hot-swapped, i.e. moved from one virtual machine platform to another or their functions switched from one virtual machine platform to another, in order to avert a disruption when CPPs emerge within virtual machines, virtual machine platforms or hardware platforms.  A virtual machine could easily be diagnosed for CPPs because it would have no proprietary components and would be completely transparent to the owners/users.  Its source code would be completely available to a weapon systems engineer.

Within the same set-up of hardware platforms, virtual machine platforms and virtual machines, there would be peer hardware platforms, peer virtual machine platforms and peer virtual machines and systems for determining the level of efficiency of the same.  When chronics of interest are detected on board a mission critical weapon system peer, there would be a seamless hot-swap of functionality to more efficient components, to ensure that mission critical operations are least affected by chronics.  The switching mechanisms would be known as swap stations/systems controlled by confidence modules described in [1].

The virtual machines/components could run on virtual machine platforms but not necessarily the ones currently in the market, as they are likely to be generic and not sufficiently specific to eliminate a wide range of easily avoidable CPPs.

Once the issue of relationships between the hardware platform and virtual machine platform is subcontracted out, the relationships between the same shall be a non-issue to weapons systems engineers and their focus shall be the relationships between the virtual machine platform that is clearly known to them and the mission critical virtual machine systems that reside on them as the control base for lethal autonomous weapons.  These lethal autonomous weapon virtual systems shall be 100% mapped out and analyzed by monitoring systems in the integrity modules that pass on data to the confidence module.

The integrity and confidence modules must be complex software systems but they could be several also hosted as virtual machines to allow for their hot-swapping should the need arise.  This brings into focus the "Gremlins Paradox" miniaturization to actualize such a kind of system shall occur but ultimately new problems shall arise and compromises shall be necessary.  Techniques e.g. rule-based, machine learning, statistical or visualization methods of identification of CPPs described in [2] and related publications shall convey their data streams to the integrity module which shall make the determination if or when hot-swapping between components should be commanded by the confidence modules that receive their output. A schematic of such a set-up is rendered in this paper [page 11] but the source code and/or pseudo-code are not revealed due to the fact that they shall be of a complex and context specific nature.

While a hot-swap occurs a black listed mission critical system could be healed by way of rebooting and reconfiguration, after which it shall be autonomously tested, before it is re-certified and brought back online into the mission critical realm as an available component [2: pages 48, 49], [9: page 7], and as a marked improvement on the concept of [1: page 20].

An interesting observation is that [9] embraces the concept of experimentation on a virtual machine environment, but since it utilizes fault injection techniques, the virtual machine concept in the that researcher's perspective, may have been a matter of convenience rather than long term practicality. With the confidence module, hot-swapping concept the lethal autonomous weapon shall only be run as a virtual machine on a platform/environment that is functioning within known parameters.

In simple terms CPPs data is generated by the specific instrumentation and forwarded to the integrity module, the integrity module makes a determination as to the validity of using the data from a system that has faulty operations or is functioning well. This information is then forwarded to a confidence module that selects which specific Mission Critical System to bring online via Hot Swapping.

Given that all the techniques and technologies describe exist, but not in immediate implementation, it shall take a short time to actualize the extensive research and solutions/tools that deal with CPPs in the realm of lethal autonomy. For purposes of clarity it should be noted that the confidence modules shall mediate overall supply and demand issues between mission critical systems on-board a lethal autonomous weapon depending upon data-streams received or not received from integrity modules.

**Schematics of the Protocol**

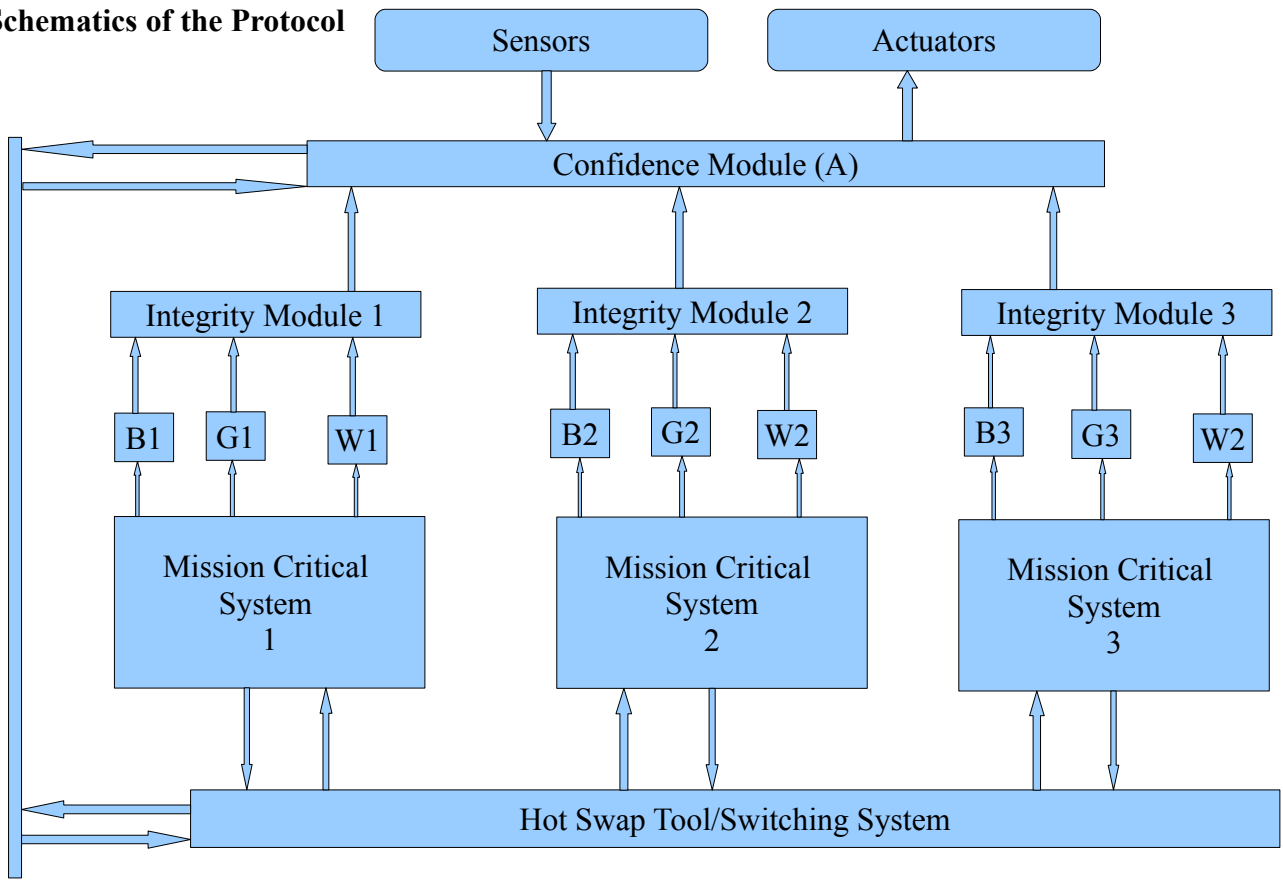

Figure Y : Schematics of the Protocol

**Layered Schematic of a Mission Critical System for a Lethal Autonomous Weapon System**

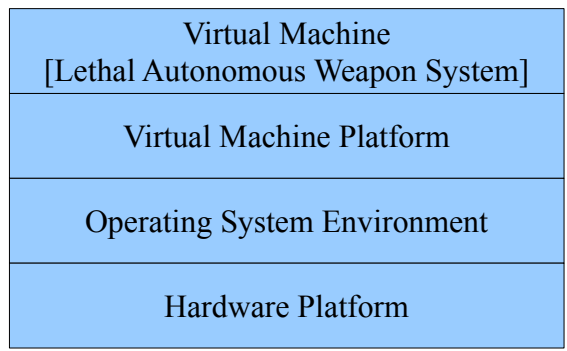

Figure K : Layered Schematic of a Mission Critical System of the type in Figure Y

Upon review, it should be clear that Figure Y above is a converged variant of Figure A in [1 : page 43] and Figure 3 in [1: page 21], an effort can be made at deducing a more converged version of the two, but that is not an issue scoped for this paper. The arrows are indicative of the data-streams/flows within the system. The letters G stands for the Gray Box Diagnostic Systems, the letters W stands for the White Box Diagnostic Systems while the letters B stand for Black Box Diagnostic Systems. These are the systems for CPPs discovery.

There could be a partial substitution of Gs and a complete substitution of Bs in the system with the Glass Box Diagnostic System, should a procuring entity have sufficient financial leverage to demand for and obtain all relevant black box parameters, in addition to knowing how to exploit the same. Though the Gs, Ws, Bs cover a wide range of systems for diagnostics of CPPs no effort has been made to select or to promote any specific mode of the same at this level of architecture.

## 4 Conclusion

The Department of Defense of the USA and its allies are usually interested in almost "magic" type technologies, that cause technological surprise against adversarial military forces. If there was a military commander who could throw a wooden staff on to the battle-space and it turns into a viper that devours adversarial vipers the better. [2] and [9] do not meet this maxim of causing technological surprise, to the extent that the solutions described therein cannot be deployed on-board unmanned lethal autonomous weapon systems without human intervention. This paper offers some vivid insights into how such a feat should be achieved using current off-the-shelf technologies.

Though several references are made to [1] the assumption is that the hot-swap protocol proposed via this paper can be implemented in other similar contexts that are not yet publicly revealed. Hopefully lethal autonomous weapons that a fielded shall be error free or error minimal, and the CPPs shall not become a prevalent justification for collateral damage.

In order to effect the concepts in this paper, it is foreseeable that an effort has to be made towards coding already existing CPPs diagnostic systems into compliance. By use of the schematic of the protocol it is possible to quantify the effort and to deliver a solution within a sensible and clearly illustrated framework.

**References**

1. Nyagudi, Nyagudi Musandu (2014): Post-Westgate SWAT : C4ISTAR Architectural Framework for Autonomous Network Integrated Multifaceted Warfighting Solutions Version 1.0" : A Peer-Reviewed Monograph. Figshare. http://dx.doi.org/10.6084/m9.figshare.899737 Pages 20, 21, 22

2. Kavulya, Soila P. (2013 May)Doctoral Dissertation : **Automated Diagnosis of Chronic Performance Problems in Production Systems,** CMU-PDL-13-109, Submitted in partial fulfillment of the requirements for the degree of Doctor of Philosophy in Electrical and Computer Engineering, Carnegie Mellon University - Carnegie Mellon University, Pittsburgh, PA, USA
https://web.archive.org/web/20140905031846/http://users.ece.cmu.edu/~spertet/papers/soila_thesis.pdf

3. ICRC(International Committee of the Red Cross), "New Weapons", Protocol Additional to the Geneva Conventions of 12 August 1949, and relating to the Protection of Victims of International Armed Conflicts (Protocol I), 8 June 1977. - Article 36 of Additional Protocol I
https://web.archive.org/web/20140606220522/http://www.icrc.org/ihl/WebART/470-750045?OpenDocument

4. ICRC(International Committee of the Red Cross), "New Weapons" - Commentary of 1987, Protocol Additional to the Geneva Conventions of 12 August 1949, and relating to the Protection of Victims of International Armed Conflicts (Protocol I), 8 June 1977. - Article 36 of Additional Protocol I - Accessed on 5th September, 2014 at 7:36 GMT not yet available on web.archive.org
http://www.icrc.org/applic/ihl/ihl.nsf/Comment.xsp?viewComments=LookUpCOMART&articleUNID=FEB84E9C01DDC926C12563CD0051DAF7

5. Arroyo, Carlos,R et al (2013), "Quantum interference effects at room temperature in OPV based single-molecule junctions", 13th Trends in Nanotechnology Conference [ Nano Express ], Nanoscale Research Letters 2013, 8 : 234; DOI : 10.1186/1556-276X-8-234 , published :16th May 2013

6. Kopp, Dr. Carlo (2009 August). "Soviet/Russian Guided Bombs : Technical Report APA-TR-2009-0806", AIR POWER AUSTRALIA – Australia's Independent Defence Think Tank, Last Updated : Mon Jan 27 11:18:09 UTC 2014
https://web.archive.org/web/20140802040221/http://ausairpower.net/APA-Rus-GBU.html

7. Dumond, John; Eden, Rick; Folkeson, John (1994), "RAND: Weapon System Sustainment Management – A Concept for Revoluntionizing Army Logistics System – A Documented Briefing", Arroyo Center – RAND, Research done for United States Army under Contract No. MDA 903-91-C-0006, ISBN : 0-8330-1479-X, RAND, CA, USA
8. Burrell, James (2003 April), "Disruptive Effects of Electromagnetic Interference on Communication and Electronic Systems", Research Project report submission for the partial fulfillment of the requirements for the degree of Master of Science in Telecommunications – George Mason University, Advisor : Dr. Jeremy Allnutt, Director of the M.S, Telecom Program, School of Information Technology and Engineering – George Mason University, USA
9. Bare, Keith A. (2009 September), "**CPU Performance Counter-Based Problem Diagnosis for Software Systems**", Thesis submitted in partial fulfillment of the requirements for the degree of Master of Science, Thesis Committee : Priya Narasimhan, Gregory R. Gauger – Chair, School of Computer Science, Carnegie Mellon University, PA, USA
10. Nyagudi, Nyagudi Musandu (2014): Tackling the "Gremlins Paradox": Autonomous Hot Swap Healing Protocol for Chronic Performance Problem Aversion in On-board Mission Critical Systems on Lethal Autonomous Weapons. fig**share**. http://dx.doi.org/10.6084/m9.figshare.1170151

Nyagudi Musandu Nyagudi is an Independent Researcher in Nairobi, Kenya. He has experience in teaching military and security courses at professional and higher education level and has studied Informatics at postgraduate level, he has several internationally acknowledged publications - with about 20years of work experience in Information and Communications Technology as a freelance systems consultant.